\documentclass[fleqn,usenatbib]{mnras}
\usepackage{amsmath}
\usepackage{graphicx}
\usepackage{amssymb,txfonts}
\usepackage{xspace}

\newcommand{\msun}{{\rm M}_{\sun}}
\newcommand{\rsun}{{\rm R}_{\sun}}

\defcitealias{ps72}{PS72}
\newcommand{\pacz}{{\citetalias{ps72}}\xspace}
\defcitealias{ls75}{LS75}
\newcommand{\ls}{{\citetalias{ls75}}\xspace}

\title[The flow through the $L_1$ point]{A comment on the properties of the matter flow through the first Langrangian point}

\author[J. Zi\'o{\l}kowski \& A. A. Zdziarski]{Janusz Zi\'o{\l}kowski\thanks{E-mail: jz@camk.edu.pl, aaz@camk.edu.pl} and Andrzej A. Zdziarski$^{\textcolor{blue}{\star}}$\\
Nicolaus Copernicus Astronomical Center, Polish Academy of Sciences,
Bartycka 18, PL-00-716 Warszawa, Poland}

\pubyear{2021}
\begin{document}


\pagerange{\pageref{firstpage}--\pageref{lastpage}}

\maketitle

\label{firstpage}

\begin{abstract}
We analyse properties of the mass outflow from the Roche-lobe filling component of a semi-detached binary system. We follow the approaches published by Paczy{\'n}ski \& Sienkiewicz and by Lubow \& Shu, which we compare with other simplified approaches. We find that the density of the flow at $L_1$ is orders of magnitude lower than the density on the same equipotential but away from $L_1$. Furthermore, the effective cross section of the flow, after averaging over its profile of the momentum density, is much lower than some published estimates done without accounting for the averaging. Thus, the use of some simplified formulae for the density and the flow cross section can lead to overestimates of the accretion rate and of the mass contained in the $L_1$ regions by very large factors unless they are supported by simultaneous integrations of the equations of stellar structure for the outer layers of the donor. \end{abstract}

\section{Introduction}
\label{intro}

The knowledge of the mass transfer rate from the donor to accretor, $\dot M$, in accreting stellar systems is of major importance for their understanding. By comparing the transfer rate with the accretion rate we can assess whether the mass transfer is conservative or associated with mass loss, as well as examine an effect of irradiation on the mass flow. The transfer rate then determines the past and future evolution of accreting systems. Observationally, $\beta \dot M$ can be estimated from a long-term average of the luminosity in an X-ray binary, e.g., \citet*{coriat12}, \citet*{zzm19}, where $\beta\leq 1$ is the fraction of the transferred mass accreted. However, for understanding of many aspects of these systems, it is important to estimate $\dot M$ theoretically, and compare the two estimates. In semi-detached binaries, such estimates can be done based on considering the evolutionary changes of the stellar radius compared to the changes of the Roche lobe, e.g., \citet*{Webbink83}. If the mass transfer is driven by nuclear expansion of the donor, $\dot M$ is connected to the time scale of this expansion, see equation (4.16) of \citet*{fkr02}. Using both approaches, we can relatively reliably determine the rate of the mass transfer, $\dot M$.

On the other hand, sometimes it is useful to consider the dynamics
of the flow through the vicinity of the inner Lagrangian point of
the Roche lobe, $L_1$. This is the case if we need to consider the
conditions near the $L_1$ point, e.g., in order to investigate the
effects of irradiation. Such analysis is helped by approximate formulae relating the radius excess, $\Delta R$, i.e., the difference between the stellar radius and the radius of the Roche lobe, to the rate of the mass transfer $\dot M$ (\citealt{ps72}, hereafer \pacz, who used in part unpublished results of
\citealt{jed69}; \citealt{sav78, sav79,Savonije83}). On the other hand, \citet{ls75}, hereafter \ls, related the physical conditions at $L_1$ to the sound speed and the binary parameters only. Their analysis has been widely applied in several contexts, see, e.g., \citet*{Lai10}.

The accuracy of these formulae remains an open question. Does the rate of the mass transfer can be obtained from either the radius excess or the sound speed? We would like to stress that the mechanism responsible for the determination of $\dot M$ is the internal evolutionary engine of the donor. The structure of the outer stellar layers (including the vicinity of $L_1$) just adjusts to carry out whatever $\dot M$ is set by the evolutionary engine. This includes both the radius excess and the gas density at $L_1$. Here, we test the accuracy of the formulae of \pacz and \ls, and find their applicability is not universal.

\section{Estimates of the flow parameters}
\label{estimates}

\subsection{The approach of \pacz}
\label{ps72_eqs}

Here, we recall and extend some of the results of \pacz and compare
them to other expressions in literature. We use the coordinate
system in the co-rotating frame as shown by \pacz in their fig.\ A1. The
$x$ axis connects the stellar centres, and it is given in units of
the semi-major axis (equal to the separation between the stars, $A$,
for a circular orbit). The orbital motion is neglected, and thus the
system has an axial symmetry around $x$, and the $y$ axis is
perpendicular to $x$ in any direction. We will also neglect a small
difference between the cross sections at $x=0$ of the star
overflowing its Roche lobe, $y_{\rm f}$ in the notation of \pacz,
and of the equipotential surface which the star fills, $y_{\rm s}$.
This results in a factor of slightly less than unity by which the
true $\dot M$ should be multiplied (see table A1 in \pacz).

The dimensionless radius of the flow cross section at $x=0$ is then
given by (using equation A15 of \pacz at $y=0$)
\begin{equation}
y_{\rm s}\approx \left(\frac{2 A}{R} \frac{\Delta R}{R}\right)^{1/2}
\frac{\mu^{3/4}(1-\mu)^{1/4}}{1+2\sqrt{\mu(1-\mu)}}, \label{ys}
\end{equation}
where $R$ is the donor radius, $\Delta R$ is the radius excess,
i.e., the difference between the stellar radius and the radius of
the Roche lobe, $\mu\equiv M/(M+M_{\rm X})$, $M$ and $M_{\rm X}$ are
the donor and accretor masses, respectively, $A$ is given by the
Kepler law,
\begin{equation}
A^3=\frac{P^2 G M}{4\upi^2 \mu}, \label{Kepler}
\end{equation}
and $P$ is the binary period. Equation (\ref{ys}) is based on purely geometrical considerations and as such is not subject to serious errors (other than neglecting the orbital motion and approximating $y_{\rm f}$ by $y_{\rm s}$). 
The total cross section of the flow at $L_1$ is given by $\Sigma=\upi (A y_{\rm s})^2$. 

Then, \pacz assume that pressure, $p$, is given by the polytropic relation,
\begin{equation}
p=K\rho^{1+1/n}, \label{poly}
\end{equation}
where $K$ is a constant. With that assumption, they find the
distribution of the momentum density, $\rho v$, as a function of the
distance from the $L_1$ point along the $y$ axis,
\begin{equation}
\rho v \approx \begin{cases}C (y_{\rm s}^2-y^2)^{n+1/2}, &y\leq
y_s;\cr 0, &y> y_s,\cr
\end{cases} \label{rov}
\end{equation}
where
\begin{align}
&C=\left(\frac{-\Omega_0 h}{n+1/2}\right)^{n+1/2}\left[K(1+1/n)\right]^{-n}, \label{C}\\
&-\Omega_{0}= \frac{G M}{A\mu} \left[1+2\sqrt{\mu(1-\mu)}\right]
\label{C1}\\
&h=\frac{1+2\sqrt{\mu(1-\mu)}}{2\sqrt{\mu(1-\mu)}}. \label{C2}
\end{align}
We note here that $\rho v$ decreases quickly with the distance along
$y$ away from $L_1$, and its average value is given by
\begin{align}
&\langle\rho v\rangle = \frac{2}{2 n+3}\rho_{\rm L_1} v_{\rm L_1},
\label{average}\\
&\rho_{\rm L_1} v_{\rm L_1}=\left[\frac{\Delta R}{R} \frac{G
M}{R(1/2+ n)}\right]^{1/2+n}\left[K(1+1/n)\right]^{-n}, \label{rhov}
\end{align}
where the subscript $L_1$ denotes a quantity measured at $L_1$.

We can also write in general
\begin{equation}
\dot M=\Sigma_{\rm eff}\rho_{\rm L_1} v_{\rm L_1}, \label{Mdot_srv}
\end{equation}
where $\Sigma_{\rm eff}$ is the effective cross section taking into account the averaging of equation (\ref{average}),
\begin{equation}
\Sigma_{\rm eff}=\frac{\Delta R}{R}\frac{G M}{R}
\frac{P^2\sqrt{\mu(1-\mu)}}{\upi (2n+3)
\left[1+2\sqrt{\mu(1-\mu)}\right]^2}. \label{Sigma}
\end{equation}
If $M\lesssim 0.6 M_{\rm X}$, we can use the approximation to the
donor's Roche-lobe radius by \citet{Paczynski67}, $R\approx (2 G
M)^{1/3} (P/9\upi)^{2/3}$, which yields
\begin{equation}
\Sigma_{\rm eff}=R\Delta R \frac{3^4\upi \sqrt{\mu(1-\mu)}}{2 (2n+3)
\left[1+2\sqrt{\mu(1-\mu)}\right]^2}. \label{SigmaP}
\end{equation}
Thus $\Sigma_{\rm eff}\sim R\Delta R$ (as noted by \citealt{sav79}),
which is much larger than a very simple, but incorrect, estimate of
$\sim\! (\Delta R)^2$. The $\Sigma_{\rm eff}$ of equation
(\ref{Sigma}) can be compared with an estimate of $\Sigma$ in
\citet*{zwg07}, which is an approximation to a formula of
\citet{Savonije83},
\begin{equation}
\Sigma_0\approx \frac{\Delta R}{R}\frac{G M}{R} \frac{P^2}{2\upi}.
\label{Sigma0}
\end{equation}
This does not include the averaging and mass-ratio terms, and is
usually an overestimate.

Using $\dot M=\Sigma_{\rm eff}\rho_{\rm L_1} v_{\rm L_1}$ together
with equations (\ref{rhov}) and (\ref{Sigma}), we obtain
\begin{align}
&\dot M\approx \left(\frac{\Delta R}{R} \frac{G M}{R}\right)^{n+3/2}
\frac{P^2}{K^n}\frac{\sqrt{\mu(1-\mu)}}{\left[1+2\sqrt{\mu(1-\mu)}\right]^2}
C_n,
\label{Mdot}\\
&C_n=\frac{1}{\upi (2n+3)(n+1/2)^{n+1/2}(1+1/n)^n},
\end{align}
which is in complete agreement with equations (A21), (A22) and (A17)
of \pacz (who expressed $\dot M$ in terms of $A$ instead of $P$).
Note that the first and second term in equation (\ref{Mdot}) are
given by very large and very small numbers, respectively, especially
for a high $n$, and a care is needed in their numerical calculation.

We note that \citet{sav78} repeated the analysis of \pacz in a more
precise way. In particular, he took into account the orbital motion
of the matter in the binary system. His formulae are more
complicated, but the general results are similar to those of \pacz.
In particular, he obtained the same functional form of the relations
given here by equations (\ref{rov}) and (\ref{Mdot}); specifically,
with the same power exponents.

Equation (\ref{Mdot}) can give us a reasonably accurate estimate of
$\Delta R$  if we know $\dot M$. This formula takes into account the
physical state of the outflowing matter through the polytrope
constant, $K$, and the polytropic exponent, $n$. Thus, in order to
determine the values of these parameters we need a model of the
outer layers of the donor. Importantly, we definitely need these
values at the $L_1$ point rather than at the depth $\Delta R$ below
the surface of the star far from $L_1$ (as practised by some
simplified formulae users). Also, if we use directly equation
(\ref{Mdot_srv}) with estimates of $\Sigma$, we need the values of density and
pressure at $L_1$ rather than away from it at the depth $\Delta R$.

Furthermore, we need an estimate of the velocity of the flow. The
most natural assumption appears that the velocity of the matter
passing through the $x=0$ plane is similar to the sonic velocity,
$c_{\rm s}$. Arguments in favour of that were given by \ls
and \citet{sav78}. If we assume that the matter crossing the $x=0$
plane has at each point the velocity equal to the speed of sound,
then for a polytrope
\begin{equation}
v^2 = c_{\rm s}^2=(1+1/n)K \rho^{1/n}. \label{cs}
\end{equation}
We can then obtain the density profile at $x=0$ with equation
(\ref{rov}) to be given by
\begin{equation}
\rho(y)=\frac{C^{\frac{2n}{2n+1}} \left(y_{\rm s}^2-y^2\right)^n }
{\left[ K(1+1/n)\right]^{\frac{n}{2n+1}} },\quad y\leq y_{\rm s},
\label{rho}
\end{equation}
and $\rho_{\rm L_1}=\rho(0)$.

\subsection{The approach of \ls}
\label{ls75_eqs}

\ls used a different approach than \pacz. Their description of the flow
through $L_1$ point is more physical and covers much larger range of
effects than that of \pacz. On the other hand, their formulae are less
quantitative. They show that the dimensional flow radius is
\begin{equation}
r_{\rm f} \approx c_{\rm s}/\Omega \label{rLS}
\end{equation}
where $c_{\rm s}$ is the gas sound speed at $L_1$ and $\Omega=2\upi/P$ is the binary orbital frequency. By comparing with equation (\ref{ys}), we see that the two become similar (apart from the $\mu$-dependent term of the order of unity) if $\Delta R\sim H_*$, where $H_*$ is the stellar scale height (since it is defined by $c_{\rm s}^2=(H_*/R)(G M/R)$. We note that this fixes $\Delta R$ independently of the rate of the mass outflow and of the physical state of the outflowing matter (with the exception of its temperature), which does not seem to be generally valid. Still, at least in some cases (Section \ref{GX}) it gives results that are surprisingly close to the results obtained using eq. (\ref{rhov}) and assuming sonic velocity of gas at $L_1$.

As a consequence of equation (\ref{rLS}), \ls further argue that
\begin{equation}
\rho_{\rm L_1}\approx \dot M \frac{4\upi^2}{P^2 c_{\rm s}^3}. \label{rho_L1_LS}
\end{equation}
The above equation is presented in the discussion of their equation (11). Using equation (\ref{Mdot_srv}) with $v_{\rm L_1}=c_{\rm s}$, we have
\begin{equation}
\Sigma_{\rm eff} \approx r_{\rm f}^2= c_{\rm s}^2 \frac{P^2}{4\upi^2}.
\label{Sigma LS}
\end{equation}
We note that the same formulae (with the accuracy to a factor of 
$\upi$) are used by \citet{fkr02} (p.\ 352).

Also, \ls estimated the ratio of the density at the Roche equipotential inside the star far from $L_1$, which we will denote as $\rho_0$, to that at $L_1$, as
\begin{equation}
\frac{\rho_0}{\rho_{\rm L_1}} \sim  \frac{\Omega A}{c_{\rm s}}
\label{rho_ratio}
\end{equation}
(see discussion after their equation 64). 

\section{Comparison with a model of GX 339--4}
\label{GX}

As an example, we consider a model of the binary system GX 339--4.
This system accretes from an evolved low-mass star onto a black
hole, and its binary period is $P = 1.7587$\,d. Specifically, we
consider a model slightly modified with respect to the evolutionary
model D obtained by \citet{zzm19}, and assume $\dot M \approx 6.1
\times 10^{16}$\,g\,s$^{-1}$. This $\dot M$ is equal to the value
obtained from the evolutionary model by matching the rate of
Roche-lobe expansion to the stellar expansion. In that model, we
have $M = 1 \msun$, $R=2.83\rsun$, the accretor mass of $M_{\rm X} =
8 \msun$, giving $\mu=1/9$, and the radius excess of $\Delta R \approx
1.22\times 10^9$\,cm (a slightly corrected value, following from
equation \ref{Mdot}). The stellar density at the depth $\Delta R$
at the equipotential describing the Roche lobe away from
$L_1$ was found to be $\rho_0=3.1\times 10^{-6}$\,g\,cm$^{-3}$ and
the temperature was $T_0=1.56 \times 10^4$\,K. The local value of
the polytropic exponent, $n$, estimated at that depth, is $n= 6.43$.
The matter there is almost fully ionized, and $X=0.74$ and $Z=0.014$
was assumed, which implies the mean molecular weight of 0.598.
Using equation (\ref{poly}) and assuming ideal gas we obtain the
entropy parameter of $K\approx 1.55 \times 10^{13}$ (cgs), and the
sound speed of $c_{\rm s0}\approx 1.6\times 10^6$\,cm\,s$^{-1}$.
These values of $K$ and $n$ are then assumed to apply to the flow
through $L_1$ and be constant within it.

For the above parameters, we can find the velocity and density at
$L_1$ from equations (\ref{cs}) and (\ref{rho}), respectively. They
are $v_{\rm L_1}\approx 7.8\times 10^5$\,cm\,s$^{-1}$ and $\rho_{\rm
L_1}\approx 3.4\times 10^{-10}$\,g\,cm$^{-3}$. Thus, the
density at $L_1$ is as much as $\approx 10^4$ times smaller than the stellar
density at the depth $\Delta R$ (while the sound speed is by a
factor of two smaller).

The geometrical cross section of the flow can be calculated using equation (\ref{ys}). We get $y_{\rm s}\approx 0.027$, which translates into the flow radius of $2.67\times 10^{10}$\,cm and the total cross section area of $2.24\times 10^{21}$\,cm$^2$. We note that either $\rho v$ decreases fast with increasing $y$, especially for a high value of $n$, which decrease is taken into account in our formulae for either $\langle \rho v\rangle$ or $\Sigma_{\rm eff}$, equations (\ref{average}), (\ref{Sigma}), respectively. For the example considered here, $\rho v$ drops to $10^{-3}$ of its value at the point $L_1$ at $y\approx 0.0215$.

Now, let us compare the values of different parameters of the flow
obtained with the help of different formulae used in literature.

We shall start with the geometrical cross section of the flow. As
was shown above, the analysis of \pacz leads to the radius of the
flow equal $2.67\times 10^{10}$\,cm. Using estimate of
\ls given by equation (\ref{rLS}) we get $r_{\rm f} \approx 1.89
\times 10^{10}$\,cm, i.e., a value smaller by a factor of only $\approx$1.4. Taking into account the approximate character of both
estimates, the agreement is excellent. We have to remember,
however, that using equation (\ref{rLS}) requires the knowledge
of the sound speed at the $L_1$ point. In our case, this
knowledge was based on the analysis following the approach of \pacz
and on the simultaneous integration of the equations of stellar
structure for the outer layers of the donor.

Now, let us compare the values of the flow effective (as opposed to
the geometrical one) cross section area obtained with the different
formulae listed in Section \ref{estimates}. The value following from
the analysis of \pacz is given by equation (\ref{Sigma}). For the
considered example, $\Sigma_{\rm eff}\approx 2.3\times
10^{20}$\,cm$^2$. Then, the approximation neglecting the averaging of the
flow over the cross section proposed by \citet{zwg07}, of equation
(\ref{Sigma0}), yields $\Sigma_0\approx 1.5\times 10^{22}$\,cm$^2$,
i.e., almost two orders of magnitude too much. The estimate by
\ls, as given by our equation (\ref{Sigma LS}), leads to the value
$\Sigma_0\approx 3.56\times 10^{20}$\,cm$^2$. This value is only by
a factor $\approx$1.5 greater than the value obtained by us with equation (\ref{Sigma}), which again means that the agreement between the two approaches is excellent. Still, it requires the knowledge of the sound speed at $L_1$.

We next compare the values of the the density of the gas at $L_1$ obtained with the different formulae listed in Section \ref{estimates}. The value following from the analysis of \pacz is given by equation (\ref{rho}), which in our case leads to $\rho_{\rm L_1}\approx 3.4\times 10^{-10}$\,g\,cm$^{-3}$. The estimate by \ls, as given by our equation (\ref{rho_L1_LS}), leads to $\rho_{\rm L_1}\approx 2.20\times 10^{-10}$\,g\,cm$^{-3}$. This is by a factor of only $\approx$1.5 smaller than the value obtained by us with equation (\ref{rho}). This is the same factor 1.5 as found by us while comparing the different estimates of the effective cross section, which identity is a consequence of the equivalence of equations (\ref{rho_L1_LS}) and (\ref{Sigma LS}).

The value of the density of the gas at $L_1$ is
certainly one of the important parameters of the flow. The fact that
the values of this parameter obtained with two different approaches
agree so well is certainly encouraging. It also supports the claim
that this estimate is not far from the true value.

Discussing the approximate formulae present in the literature, we
should devote attention to popular formulae equivalent to
our equation (\ref{Mdot_srv}) but used in improper way, As an
example we may recall equation (9) of \citet{zwg07}. Those
authors use equation (\ref{Mdot_srv}) but they replace the effective cross
section area with the geometrical one and the density of the gas at
$L_1$ with the density on the Roche-lobe equipotential
away from $L_1$. Since, as we have seen, the effective cross section
area might be overestimated this way by two orders of magnitude and
the density of the gas even by four orders of magnitude, such
use of equation (\ref{Mdot_srv}) might lead to an error
reaching even six orders of magnitude. It is therefore crucial, while
using formulae of this type, to use the proper value of the gas
density.

Another example may be found in \citet{zzm19}. While those authors
properly estimated $\dot M$ from their evolutionary model, they
still used the above simplified treatment to estimate the mass
stored within the $L_1$ flow, and thus the emptying time of that
region, which was then used to estimate the variability time scale
related to changing conditions at $L_1$. This lead to a major
overestimate of the expected variability time scale. Using the
proper description of the flow, we can calculate the mass column
density, $\sigma$, perpendicular to the flow (i.e., the stored in a
1\,cm thick slice around $L_1$). Integrating equation (\ref{rho}),
we obtain
\begin{equation}
\sigma=\upi A^2 \rho_{\rm L_1}\frac{y_{\rm s}^2}{n+1}. \label{sigma}
\end{equation}
For the considered example, $y_{\rm s}\approx 0.027$, and
$\sigma\approx 8.4\times 10^{10}$\,g\,cm$^{-1}$. On the other hand,
\citet{zzm19} used $\sigma =\rho_0 \Sigma_0$, which is $4.8\times
10^{16}$\,g\,cm$^{-1}$, a factor of $\approx 5\times 10^5$ too much.
This then lead to a corresponding overestimate of the emptying time,
$\Delta t$. Those authors also used the $\dot M$ following from
their evolutionary model and approximated the length of that region
as $\Delta R$, yielding $\Delta t\approx 24$\,yr. If we make the
same assumptions as them but use the correct $\sigma$, we obtain
$\Delta t\approx 1300$\,s. Even if we change the assumption about
the length of the $L_1$ element made in \citet{zzm19} to a more
realistic value, we can increase the above $\Delta t$ by at most two
orders of magnitude. Thus, the conclusion of them that $\Delta t$
can account for the observed long-term changes of the accretion rate
average over outburst is incorrect, and another mechanism is needed.

Finally, we should comment on the discrepancy between the
the density of the gas at the Roche equipotential inside the star
far from $L_1$ obtained in our calculations and the density obtained
with the prescription of \ls, equation (\ref{rho_ratio}). Using it, we find the ratio of $\approx$52. As we saw in our analysis above, that ratio was about four orders of magnitude ($3.1\times 10^{-6}$ vs $3.4\times
10^{-10}$\,g\,cm$^{-3}$). This is a serious discrepancy, by more
than two orders of magnitude. However, we believe that our estimate
is closer to the real situation than that of \ls. The
reason is the following one. We started with the determination of
the radius excess, $\Delta R$. We used for that
purpose equation (\ref{Mdot}). Fortunately, the sensitivity of $\Delta
R$ to $\dot M$ is very weak. As demonstrated by \citet{zzm19} (see their
fig.\ 3), increasing $\dot M$ by four orders of magnitude leads to
increase of $\Delta R$ by a factor of only two. Since we have a
relatively reliable estimate of $\dot M$ based on evolutionary
considerations, we may trust that $\Delta R$ is determined with
rather good precision. Knowing the depth of the Roche
equipotential below the stellar surface, we may integrate the stellar
structure equations (including the heat transfer equation) from the
stellar surface inward. At the depth $\Delta R$ we determine, among
others, the density of the gas and the gas sound speed. This
procedure is rather straightforward and should not raise serious
doubts. On the other hand, the \ls estimate is based on
rather qualitative analysis of a very complicated flow in the outer
stellar layers.

Finally, we should stress that all the considered descriptions  are
still quite approximate. Hydrodynamic simulations are needed to
get more accurate descriptions.

\section{Conclusions}
\label{concl}

We have overviewed and extended the treatment of the flow through
$L_1$ of \pacz, and compared it with other, simplified, methods. We
have stressed that the flow density at $L_1$ can be several orders of magnitude lower than that at the same equipotential but away from $L_1$.

Furthermore, one of the approximate estimates of the flow effective
cross section used in literature overestimates the actual one by two
orders of magnitude. The other one, given by \ls, leads to
result that remains in excellent agreement with our estimate. The
same is true about the estimate of the gas density at $L_1$ given by
\ls (but to calculate it correctly we have to know the
structure of the outer layers of the donor).

However, we have found that the estimate of the ratio of the density of the
gas at the Roche equipotential inside the star far from $L_1$ to that at $L_1$
according to the prescription of \ls is not accurate.

The overestimates of the density and the cross section in
some formulae then can lead to a gross overestimate of the outflow
rate (calculated as a function of the parameters of the flow), as
well as of the amount of mass present in the $L_1$ flow. We should,
however, remember that outflow rate is determined by the
evolutionary engine of the donor.

Generally, we conclude that such simplified formulae can give
reasonably accurate estimates if they are supported by simultaneous
integrations of the equations of stellar structure for the outer
layers of the donor.

\section*{Acknowledgements}

We thank the referees for valuable suggestions. This research has been supported in part by the Polish National Science Center grants 2015/18/A/ST9/00746 and 2019/35/B/ST9/03944.

\section*{Data Availability}

There are no new data associated with this article.

\label{lastpage}
\end{document}